\documentclass[epj]{svjour}

\pdfoutput=1

\usepackage{graphics}
\usepackage{epsfig}
\usepackage{amssymb}
\usepackage{subfigure}
\usepackage{color}
\usepackage{pifont}
\begin{document}
\title{Non-local rheology in dense granular flows}
\subtitle{Revisiting the concept of fluidity}
\author{Mehdi Bouzid \and Adrien Izzet \and Martin Trulsson \and Eric Cl\'ement \and Philippe Claudin \and Bruno Andreotti}             
\institute{Physique et M\'ecanique des Milieux H\'et\'erog\`enes, UMR 7636 ESPCI -- CNRS -- Univ.~Paris-Diderot -- Univ.~P.M.~Curie, 10 rue Vauquelin, 75005 Paris, France.}

\date{\today}

\abstract{
The aim of this article is to discuss the concepts of non-local rheology and fluidity, recently introduced to describe dense granular flows. We review and compare various approaches based on different constitutive relations and choices for the fluidity parameter, focusing on the kinetic elasto-plastic model introduced by Bocquet et al. [Phys. Rev. Lett \textbf{103}, 036001 (2009)] for soft matter, and adapted for granular matter by Kamrin et al. [Phys. Rev. Lett. \textbf{108}, 178301 (2012)], and the gradient expansion of the local rheology $\mu(I)$ that we have proposed [Phys. Rev. Lett. \textbf{111}, 238301 (2013)]. We emphasise that, to discriminate between these approaches, one has to go beyond the predictions derived from linearisation around a uniform stress profile, such as that obtained in a simple shear cell. We argue that future tests can be based on the nature of the chosen fluidity parameter, and the related boundary conditions, as well as the hypothesis made to derive the models and the dynamical mechanisms underlying their dynamics.
}

\maketitle
%

\section{Introduction}
\label{intro}

Since \emph{non-locality} was introduced as an interpretive framework for dense granular flows \cite{AD01,R03,J06,A07}, it has become a key concept to describe the rheology of complex fluids in soft condensed matter. However, the connections between the various contributions to this subject, their similarities and possible conflicts need clarification. In particular, among the pending questions that must be answered, a fundamental and vivid issue is the possible emergence of non-locality as the signature of a dynamical phase transition \cite{GCOAB08,BCA09,GCB10,CMCB12,MC12}. This interrogation does not only concerns granular matter but should be apprehended in the more general context of amorphous solids undergoing a rigidity transition. At present, different conceptual approaches have been put forwards to describe non-locality and several non-local constitutive relations were proposed for granular matter. It is thus fair to ask whether these approaches are equivalent and to which extent for example, they are similar to phase field models \cite{AT01,AT02,VTA03,ATMC08} built on an underlying liquid/solid phase transition. Also for granular matter, shear banding and apparent ``creeping zones'' are observed which are difficult to re-conciliate with a simple local rheology \cite{O08}, and this has been the starting point of different propositions for non-local constitutive relations. For many, the elements of proof validating these approaches has often been a mere ''good fitting'' of the velocity profiles. A legitimate question is then to ask whether this is sufficient to demonstrate the validity of a particular model and moreover, what could be other more stringent tests providing essential information on the dynamical mechanisms responsible for non-locality.

Our aim here is to propose a critical discussion of the concepts of non-local rheology and fluidity in dense granular matter, based on recent progresses as well as older results. In the next section, we first review the rheology of dense granular flows, starting from the local rheology towards evidence for non-local effects and describing non-local approaches. In section~\ref{fluidity}, we discuss the concept of fluidity. Section~\ref{differences} is devoted to the differences among the non-local constitutive relations proposed for granular matter. We end the paper in section~\ref{furthertests} with a discussion on further possible tests that must be performed to better understand the mechanisms at the origin of non-local effects.

\section{On the rheology of dense granular matter}
\label{rheology}

\subsection{Rigidity transition}
When sufficiently polydisperse to avoid crystallisation, a granular packing at rest can be considered as an amorphous solid. By definition, \emph{amorphous solids} refer here to systems that may resist to a shear stress while they do not present any long range translational order at the microscopic scale, namely the grain size for granular matter. Let us consider, for clarity, an ideal rheometer in which the material is submitted to a homogeneous shear stress $\sigma$. The system behaves mechanically as a \emph{solid} if it reaches equilibrium at a finite strain $\gamma$. It is considered as \emph{elastic} if it returns to its original state, once the stress is removed and \emph{plastic}, otherwise. Conversely, the system behaves mechanically as a \emph{liquid} if it flows permanently at a finite strain rate $\dot \gamma$. The system exhibits a \emph{rigidity transition} if its dynamical behaviour switches from solid-like to liquid-like, when a control parameter crosses a threshold value. Most soft amorphous solids present a rigidity transition upon varying the shear stress, the threshold value being named the \emph{yield stress} $\sigma_y$. We hence define the \emph{yield parameter} as the ratio of the shear stress to the yield stress:
\begin{equation}
\mathcal{Y}=\frac{\sigma}{\sigma_y}.
\label{defmathcalY}
\end{equation}
We emphasise here that the existence of a yield stress is not an intrinsic material property.  It depends on the other control parameters that are kept constant during the loading. As an example, a granular material, dry or suspended in a fluid, displays a yield stress when the particle-borne pressure $P$ is imposed, whereas it does not when the volume fraction $\phi$ is imposed. This is obviously a key issue and often a source of confusion. This question was discussed recently in the context of different amorphous particulate materials \cite{TBKCCA15}. 

The rigidity transition is intimately related to the multi-stability of the energy landscape: the system has to cross energy barriers to flow. The physical nature of the mechanisms preventing irreversible plastic deformations allows to classify the soft amorphous materials and their corresponding rigidity transition as: 
\begin{itemize}
\item Entropic for glasses formed by thermal quenching \cite{EAN96,DS01,MKK09}; the rigidity transition is then called "glass transition" \cite{OT07}.
\item Enthalpic for soft elastic particles at high volume fraction \cite{PvM85,MW95,HW95,MW12,CBML03}; the free energy may result from capillarity (foam, emulsion), from electrostatics, from the particle elasticity, etc. The rigidity transition is then called elasto-plastic depinning transition \cite{F98,TLB06,HKLP10}.
\item Geometric for hard grains submitted to a confining pressure \cite{AFP13}; the rigidity transition is then called jamming transition \cite{LN98}.
\end{itemize}
Although we focus here on the last case, we will frequently discuss connections with other complex fluids in soft matter.

\subsection{Local rheology}
Following the seminal paper by GdR MiDi \cite{GDRMidi}, major improvements were obtained to provide a consistent framework to understand and model how granular matter flows. In the rigid limit, granular matter does not present any intrinsic energy scale and confining pressure $P$ thus provides the only relevant scale of energy per unit volume. As a consequence, $P$ sets the yield stress as $\sigma_y=\mu_c P$, where $\mu_c$ is the "critical" friction coefficient, which depends on microscopic material properties (e.g. packing polydispersity, inter-granular friction, shape etc...). For a real granular material, the rigidity transition is actually subcritical, a property associated with the presence of inter-granular friction, and the hysteresis of the effective friction coefficient has remained unexplained up to now. For grains of mean diameter $d$ and mass density $\rho_g$, the confining pressure also sets the time-scale $T=d/\sqrt{P/\rho_g}$ for plastic reorganizations at the granular level (microscopic time scale). Following \cite{GDRMidi,CEPRC05}, one can define the rescaled strain rate, or \emph{inertial number}, as:
\begin{equation}
I\equiv \dot \gamma T=\frac{\dot \gamma d}{\sqrt{P/\rho_g}} \, .
\end{equation}
Writing the yield parameter as
\begin{equation}
{\mathcal Y}=\frac{\sigma}{\sigma_y}=\frac{\sigma}{\mu_c P} \, ,
\end{equation}
the constitutive relation for homogeneous steady flows takes the generic form 
\begin{equation}
{\mathcal Y}=\frac{\mu(I)}{\mu_c}=1+a I^n.
\label{HB}
\end{equation}
$n=1$ for grains presenting a standard friction coefficient $\simeq 0.5$ at contact, and $n=1/2$ for frictionless grains. This relation must be complemented by a law relating the volume fraction $\phi$ to the rescaled strain rate $I$:
\begin{equation}
\phi-\phi_c =-b I^n,
\label{HBPhi}
\end{equation}
with the same phenomenological exponent $n$; $a$ and $b$ are constants that depend on the microscopic details of the system. This derivation, is simply based on dimensional analysis, in the rigid limit where the grain elasticity is irrelevant. Empirical measurements indeed show a frictional behaviour with the emergence of a yield stress. 

Relation (\ref{HB}) is well suited to investigate pressure-controlled situations. However, the same equations can entirely be recast to handle situations where $\phi$ is fixed, and then the yield stress disappears from the constitutive picture. Inverting eq.~(\ref{HBPhi}), one obtains:
\begin{eqnarray}
P&=&b^{2/n}\,\frac{ \rho \dot \gamma^2 d^2}{(\phi_c-\phi)^{2/n}},\\
\sigma&=&\mu_c b^{2/n}\left(1+\frac{a}{b} \left(\phi_c-\phi \right)\right)\frac{ \rho \dot \gamma^2 d^2}{(\phi_c-\phi)^{2/n}} \, .
\end{eqnarray}
In this representation, the shear stress clearly vanishes in the limit $\dot \gamma \to 0$, i.e. if $\phi<\phi_c$. From this derivation, we can see that, even in the limit of rigid particles, i.e. without any explicit elasticity, a granular material is compressible under shear \cite{TBCA13}. The pressure therefore requires time to establish over the size of the system. Provided this time-scale remains short compared to $\dot \gamma^{-1}$, the pressure can be considered as defined ''instantaneously'' and thus can be used as a state variable instead of $\phi$. This is indeed a central assumption to neglect compressibility (Boussinesq approximation) when using the $\mu(I)$ formulation in heterogeneous situations, : it implicitly requires that pressure is established macroscopically over a very short time scale and varies slowly in time and space.

Using this close set of constitutive relations, a quantitative agreement with numerical or experimental measurements has been reached in different configurations. In particular, for avalanche flows of glass beads on a rough inclined plane -- an important situation as the yield parameter ${\mathcal Y}$ is fixed by the inclination angle -- Pouliquen \cite{P99} has derived an effective flow rule consistent with the $\mu(I)$ rheology. This has led to a three-dimensional extension of the local rheology \cite{JFP06}, yielding the correct scaling laws characterising the chute flow geometry.

Dense granular suspensions have been shown to follow the same rheology, with $T=\eta_f/P$, where $\eta_f$ is the viscosity of the suspending fluid, to form the so-called viscous dimensionless number $T\dot\gamma$ \cite{BGP11,TAC12} --~sometimes noted $J$. More generally, the rheology obeyed by a granular material in a homogeneous steady state takes the very same form as observed for soft material presenting a yield stress. In those more general situations, the shear rate $\dot \gamma$ can then be rescaled by a plastic time-scale $T$, to form a dimensionless number akin to $I\equiv \dot \gamma T$. This kind of rheology takes the same generic form (\ref{HB}), often called a Herschel-Bulkley constitutive relation, for all these systems.

\begin{figure*}[t!]
\includegraphics[scale=.3]{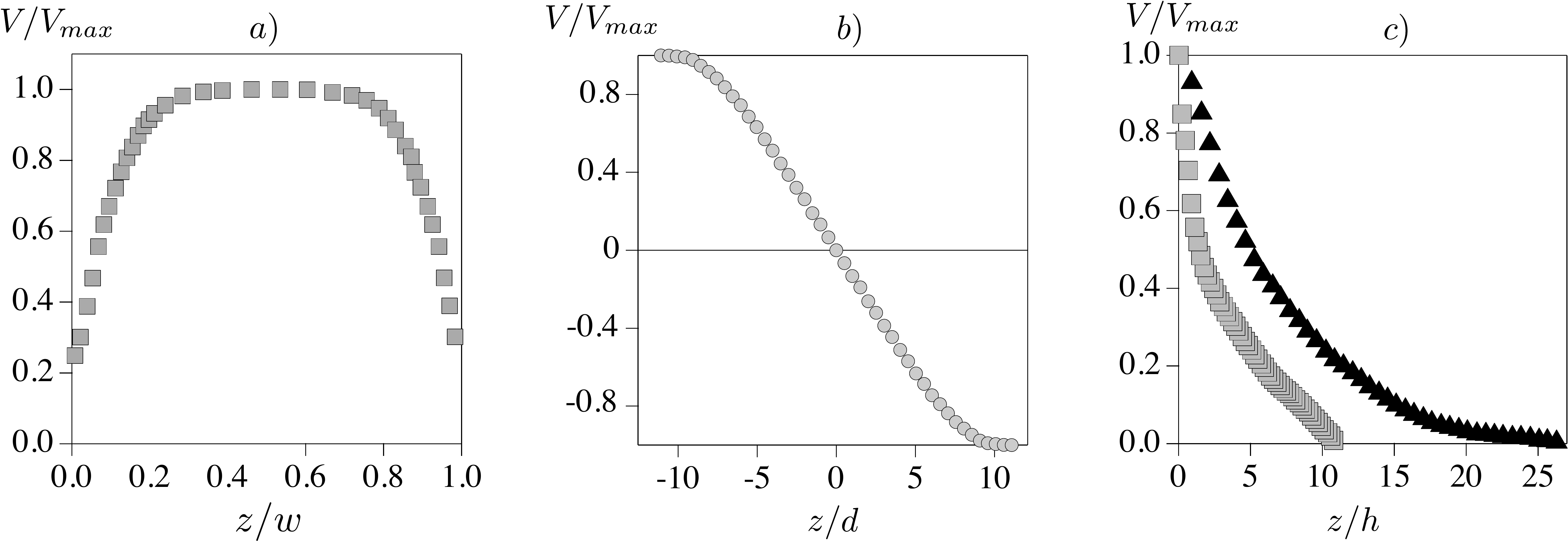}
\vspace{- 0 mm}
\caption{Experimental and numerical results for soft and granular matter displaying non-locality. (a) Velocity profile of the flow of a jammed emulsion in a micro-channel with rough surfaces, data from \cite{GCOAB08}. Local rheology would predict a plug flow. (b) Velocity profile of a granular flow in a 2D shear cell. Local rheology would predict a linear velocity profile. The numerical data are from \cite{BTCCA13}. (c) Experimental (triangles) and numerical (squares) velocity profiles of a foam flow in a Couette cell, data from \cite{KTMvH10} and \cite{BSSPBST15} respectively. Local theory would predict a localised failure at wall.}
\vspace{0 mm}
\label{Fig1}
\end{figure*}

\subsection{Failure of the local rheology}
Consider now a heterogeneous shear flow. Its rheology is said \emph{local} if the stress tensor at a given location is still a function of the shear rate at the same place. \emph{Non-locality} refers to any deviation to such a local constitutive relation. Before giving precise examples, let us discuss the choice of such a name. It has been introduced in granular material to describe the distant transmission of momentum through the granular skeleton, during collisions  \cite{AD01,R03,J06,A07}. In the limit of rigid particles, this transport of momentum is instantaneous so that a stress, which is a flux of momentum, can be induced by distant collisions. 

In hydrodynamics, the epitome of non-locality is pressure. In a simple fluid, pressure is transmitted at the speed of sound. At low mach numbers, the time needed for a pressure signal to cross the entire flow is small compared to $\dot \gamma^{-1}$. In this limit, pressure is determined by the incompressibility condition $\vec \nabla \cdot \vec v=0$, where $\vec v$ is the velocity field. Taking the divergence of the Navier-Stokes equation one therefore obtains, for a Newtonian fluid, the Laplace equation:
\begin{equation}
\vec \nabla^2 P=-\rho \vec \nabla \cdot (\vec v\cdot \vec \nabla \vec v).
\end{equation}
Pressure balances the potential part of inertial terms. The Biot-Savart equation provides an explicit solution of this equation under an integral form. The fact that pressure is a function of the whole velocity field, and not only of the local strain rate then appears explicitly. As a conclusion, there are \emph{a priori} two definitions of non-locality which are not equivalent:\\
(i) Momentum is transported over large distances on a time-scale small in comparison to $\dot \gamma^{-1}$ and to the plastic rearrangement timescale $T$.\\
(ii) The constitutive relation involves a second state variable, which is not a function of the strain rate, and whose evolution is controlled by an independent equation, typically involving a Laplacian operator.\\
For instance, the kinetic theory, which is valid for dilute and rapid granular flows, is non-local in the weak sense (ii) since it introduces an independent field representing the mean squared velocity fluctuations (the so-called granular temperature), which may control the stress tensor \cite{JS83}. However, the transmission of momentum remains perfectly local in the sense (i). 

Non-locality in the weak sense (ii) manifests itself through different properties. The first one is the evidence of a creeping flow in regions below the yield condition (${\mathcal Y}<1$) \cite{GDRMidi,KINN01,TRVLPJD03,TRD08,NDBC11}. Instead of the expected static zone (i.e. a solid), one observes an exponential spatial relaxation of the shear rate $\dot \gamma$. A second property is the fact that the yield conditions are sensitive to the system size and to the boundary conditions. In the case of granular matter, the yield stress measured on an inclined plane depends on the thickness of the deposit \cite{P99,GDRMidi}. Also, in conditions where the grains should flow according to the local rheology, jammed regions are identified below the flow \cite{MAC15}. Furthermore, for self-channelised flows close to jamming, quantitative departure from the local rheology predictions are explicitly shown \cite{DLDA06}. Finally, properties pointing on the existence of non-local effects are revealed by micro-rheology experiments. For example, the force-velocity relations assessed by an intruder plunged in the material strongly depend on the presence of a distant shear flow \cite{NZBWH10,RFP11,WH14}, or of a vibrating boundary \cite{HDKC11}. Examples of manifestation of non-locality for soft and granular matter are given in Fig.\ref{Fig1} and Fig.\ref{Fig2}.

\subsection{Reviewing non-local models}
We review here the main approaches that were put forwards to tackle these problems.

\subsubsection{Cosserat approach}
In the framework of plastic theories developed for soil mechanics, the formation of localised shear bands is sometimes apprehended via a Cosserat extension of the continuous elasto-plastic theory (see for example \cite{MV87} and refs inside). To describe the quasi-static state of deformations of a granular material, new fields are introduced that couple stresses and rotations. This theory introduces a microscopic length scale describing the range of influence of the microscopic granular rotations and provides a non-local coupling for plastic deformations over this scale. However, in spite of the fact that it may provide a useful regularisation technique for numerical computation methods, its use is often seen as limited since the issue of assessing objectively the constitutive parameters and providing consistent boundary conditions for the fields, has so far remained a short-coming of the approach.

\subsubsection{Phase field approach}
As stated above, non-locality, in the weak sense, reflects the existence of a state parameter, beyond the strain rate, determining the stress values. As this parameter measures how fluid the system is, following Derec et al. \cite{DAL01}, we will refer to it as the "fluidity" and will note it thereafter $f$. We warn the reader that in different papers, fluidity may refer to different  physical quantities. Here, we keep the name and its conceptual definition in relation to non-local rheology. Importantly, we consider that for our purpose, a relevant fluidity has to be selected on physical basis, amongst all state variables. 

From a phenomenological point of view, \emph{fluidity} plays the role of an order parameter describing the dynamical transition from solid-like to fluid-like behaviour. It was first proposed by Aranson \& Tsimring to introduce a phase field $f$ which vanishes in the static state and which tends to $1$ in the fully fluidised state \cite{AT01}. In this approach $f$ is therefore dimensionless. The overall shear stress is then formally decomposed as the sum of a solid and liquid like contributions weighted respectively by $1-f$ and $f$. Following Landau standard derivation, the order parameter $f$ is controlled by a reaction diffusion equation of the form:
\begin{equation}
T \dot f={\mathcal I}(f)+\ell^2 \vec \nabla^2 f,
\end{equation}
where ${\mathcal I}$ is a function of $f$ parametrised by the state variables and in particular by the rescaled shear rate $I$. Note that ${\mathcal I}(f)$ can be designed to reflect a subcritical, hysteretic transition from solid-like to liquid-like behaviours, as generically observed for granular matter. The microscopic time $T$ is a characteristic time for fluidization to occur and $\ell$ an elementary length scale. The diffusion coefficient of the fluidity is $\ell^2/T$. The Laplacian operator results from a gradient expansion, assuming spatial isotropy. This term is the transcription of a (weak) non-locality. As a matter of fact, in the steady state, $f$ is determined by a non-linear Laplace equation, just like pressure in hydrodynamics. 

This approach produces  an effective rheology different from $\mu(I)$ in the sense that Pouliquen's flow rule \cite{P99}, valid for avalanches of spherical beads, is not recovered. However, it yields a real solid/fluid phase transition semi-quantitatively close to what is observed for sandy grains, in particular the starting and stopping heights, flow rules and erosion/deposition waves \cite{MLAC06,AMC06,CMAA07,ATMC08}. Nevertheless, it does not reproduce creep zones close to a shear band.

\subsubsection{The elasto-plastic approach}
The approach proposed by Kamrin et al. \cite{KK12} to model granular flows is directly adapted from the Kinetic Elasto-Plastic (KEP) model introduced by Bocquet et al. \cite{BCA09} for soft matter. The key concept is the fluidity, which can be defined in an unified way as:
\begin{equation}
f=\frac{\dot \gamma}{\mathcal Y}.
\end{equation}
We will devote below an entire sub-section to this approach.

\subsubsection{Mechanically activated plastic events}
An original idea to describe non-locality has been proposed by Forterre and Pouliquen \cite{PF09}. It is based on an analogy with Eyring's transition state theory for the viscosity of liquids, where mechanical fluctuations --~introduced here as a synonymous of heterogeneities~-- would play the role of temperature in thermal systems. Plastic rearrangements occur at a rate proportional to the strain rate $\dot \gamma$. They are assumed to generate at random a new realisation of the forces on the contact network, allowing for the formation of new weak zones where the next rearrangement will occur. 

At a semi-quantitative level, this approach improves significantly the local visco-plastic approach as it proposes a physical hint for microscopic processes inducing non-locality for granular media. Moreover, it can predict dependence of the stopping angle with flow height, shear bands extension increasing with the flow rate. However, the theoretical outcomes are more difficult to quantitatively re-conciliate with the Pouliquen's flow rule for the chute flow and more importantly, it does not predict a thickness dependence of the avalanche starting height. In the non-local formulation proposed by Pouliquen and Forterre \cite{PF09}, the shear rate $\dot \gamma$ obeys an integral equation, which involves an exponential kernel, function of the stress tensor and of the distance, interpreted as a Boltzmann-like factor. Provided the fact that for granular packing, the stress fluctuations take place generically over few grain sizes, the authors assume a spatial dependence of the interaction kernel as a Lorentzian function decaying algebraically (power $-2$) over this granular size.

The relation between the shear stress and the strain rate depends non-locally on two fields i.e. on two fluidity parameters: the strain rate $\dot \gamma$ and the confining pressure $P$. The relatively fast decay of the chosen spatial kernel makes possible a long wavelength expansion of the rearrangement rate equation with respect to $\dot \gamma$ and $P$. If $P$ varies slowly at the scale of the grain size, the expansion generates at the first order a laplacian operator, like other models. More precisely, the constitutive relation presents a dependence with $\vec \nabla^2 \dot \gamma$, suggesting that the fluidity $f$ is the rate of plastic events $\dot \gamma$.

\subsubsection{The gradient expansion of the constitutive relation}
We have ourselves followed an approach which is significantly different from the previous ones, and which suggests possible candidates for the most appropriate fluidity of dense granular flows \cite{BTCCA13,PhDBouzid14}. Imagine the problem solved and a fluidity $f$ built, which vanishes in the solid phase and increases with the ability to flow. We then follow the standard vision of Maxwell rigidity transition as put forwards to understand jamming in granular matter. Due to the cooperative motion of particles along soft modes, flowing is facilitated when, at a given point it is surrounded by a more fluid zone. Conversely, the resistive stress is larger when the point is surrounded by a more solid neighbourhood. Experimentally, the effect is particularly significant close to the jamming transition, when $f$ vanishes. Therefore, one needs to define a relative fluidity, which compares the degree of fluidity at one point and in its vicinity. Assuming as before, that the influence of fluidity is statistically isotropic and results from a short range interactions between shear zones,  the relative fluidity can be defined as:
\begin{equation}
\kappa = \frac{\ell^2 \,\vec \nabla^2 f}{f},
\end{equation}
where $\ell$ is a length on the order of few grain diameters. The Laplacian is indeed the lowest order operator appearing in a systematic expansion in a functional of $f$. The relative fluidity $\kappa$ remains finite when $f$ goes to $0$. The constitutive relation can be expanded around the relation $\mathcal{Y}=\mu(I)$, valid in the homogeneous case, according to:
\begin{equation}
\mathcal{Y}=\mu(I) \chi(\kappa) \quad {\rm with} \quad \chi(\kappa) \simeq 1-\kappa +\mathcal{O}(\kappa^2).
\label{NLfriction}
\end{equation}
$\kappa$ is positive when the point considered is surrounded by a more liquid region (higher $f$). This region flows more easily than expected from the local value of $f$, and the corresponding shear stress is therefore lower. In this formulation, $\chi(\kappa)$ must thus be a decreasing function of  $\kappa$, which justifies the minus sign in front of $\kappa$ in (\ref{NLfriction}). Note that the lack of multiplicative factor in Eq.~(\ref{NLfriction}) defines $\ell$ in a univocal way.  Importantly, this phenomenological derivation does not depend on the nature of the mechanical interaction between  the shear zones; the reader may think of the analogy with the van der Waals gradient expansion of the Helmholtz free energy at a liquid-vapour interface.

There are three obvious choices for the granular fluidity. One would be to  introduce the (coarse grained) number of contacts per grain $Z$ and number of sliding contacts per grain $\zeta$. In the spirit of Maxwell rigidity transition theory, the fluidity would then be defined as the distance to isostaticity. A second possibility is to introduce the volume fraction $\phi$, and to build the fluidity as the distance $\phi_c-\phi$ to the value $\phi_c$ that the volume fraction reaches in the limit $I \to 0$. This poses two problems: first, $\phi_c$ depends on the fraction of sliding contacts; second, one would need to introduce the mass conservation equation for $\phi$ i.e. to consider explicitely the granular fluid as compressible. Finally, there is a last quantity that can play the role of the fluidity: the inertial number $I$ itself. It vanishes in the solid state and increases with the degree of fluidity. Ockham's razor --~law of parsimony~-- is obviously in favour of such a choice, as the equations are closed without involving further equations, in the dense limit $\phi\simeq \phi_c$. This does not constitute a deep scientific argument, except that simple models are better testable. 

The constitutive relation (\ref{NLfriction}) must actually be complemented by another one for the volume fraction. Assuming that $\phi$ is a local function of $I$, expressing the non-local rheology with $f=I$ with $\kappa = \ell^2 \nabla^2 I/I$, and with $f=\phi_c-\phi$ with $\kappa = \ell^2 \nabla^2 \phi/(\phi_c-\phi)$ are mathematically analogous. The only subtlety is that compressibility must be taken into account if the non-local constitutive relation is expressed in terms of $\phi$ while the flow can be considered as almost incompressible (i.e. in the Boussinesq approximation) if the non-local constitutive relation is expressed in terms of $I$. In the later case, the pressure $P$ is instantaneously determined and becomes a state variable.

To conclude this section, we rewrite the non-local rheology (\ref{NLfriction}), with the choice of $f=I=T \dot \gamma$ as a fluidity, as:
\begin{equation}
0=\frac{I {\mathcal Y}}{\mu(I)}-I+\ell^2 \vec \nabla^2 I.
\label{Mehdi}
\end{equation}
This formulation, directly derived from Eq.(\ref{NLfriction}), makes it easier to compare with the other approaches, as discussed below.

\section{The fluidity concept in soft elastic materials}
\label{fluidity}

Since fluidity was introduced to describe granular flows, it is important to review the physical basis of the recent advances made to render the complex rheology of soft matter (foams, gels, emulsions etc...) using this concept.  Many amorphous materials, including granular matter share the same phenomenology, e.g. yield stress, Herschel-Bulkley rheology. Although fluidity refers in standard rheology to the inverse of viscosity, this name has been associated with different quantities in the literature.  In the recent conceptual picture derived from soft matter, fluidity appears as a variable entering into the rheological constitutive relation of the material. Qualitatively, low fluidity means closer to a solid and large fluidity means closer to a fluid. Importantly, in those approaches fluidity dynamics obeys an auxiliary equation which sets-in its temporal evolution. This auxiliary equation reflects the microscopic or mesoscopic processes at the heart of the physics that one seeks to describe. We shortly review the propositions made in the literature to define fluidity.

\begin{figure*}[t!]
\includegraphics[scale=.3]{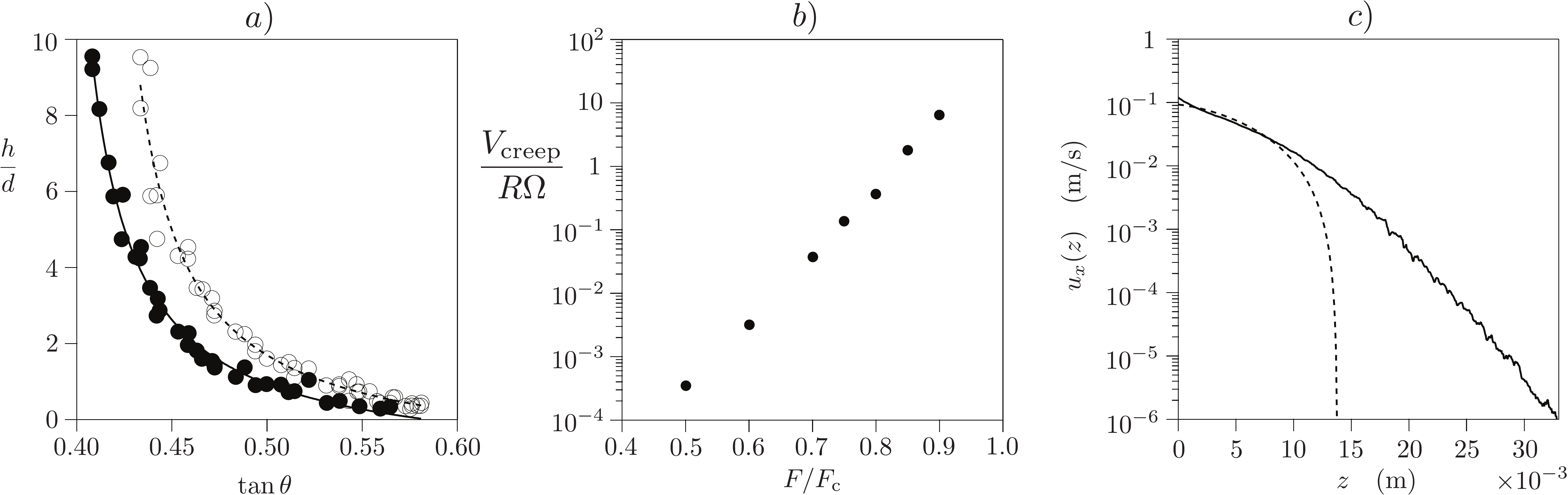}
\vspace{- 0 mm}
\caption{Evidences of Non-local behaviour for granular matter. (a) Experimental measurements of the stopping (black symbols) and starting (white symbols) heights on an inclined plane with glass beads, data from \cite{GDRMidi}. The dependence of stoppage and onset of flow with the granular height are manifestations of the non-local character of the rheology. (b) Normalised creep velocity of an intruder in a Couette cell filled with glass beads as a function of the normalised pulling force on the intruder, data from \cite{RFP11}. Data are taken in conditions where the intruder should be blocked (in the solid phase) in a local rheology picture based on a Coulomb yield criterion. (c) Experimental velocity profile of a dry granular avalanche flow in a narrow channel with frictional lateral walls. The dashed line is the prediction of the local rheology displaying a depth $z$ where flow stops whereas the granular continues to flow below this limit.}
\vspace{0 mm}
\label{Fig2}
\end{figure*}

\subsection{Fluidity and elastic deformation}
Many complex fluids are visco-elasto-plastic. It is the case of polymer melts, micelles and lamellar surfactant phases, which are subject to shear-banding, but also the case of foams and emulsions, which are yield stress fluids. In this situation, an obvious state variable controlling the rheology is the elastic strain $\epsilon$. The elastic deformation is an intrinsic property describing for example, the polymer chains extension or for a foam, the current bubble deformation state. Importantly, $\epsilon$ is a field coarse grained in space, in time, or averaged over realisations. The total strain rate $\dot \gamma$, determined from the subsequent positions of the elements, can be formally written as the sum of the elastic strain rate $\dot \epsilon$ and of a plastic contribution, equal to $\dot \gamma-\dot \epsilon$. For convenience, we keep a scalar description -- the extension to a tensorial form does not present any conceptual difficulty.

Non-locality in visco-elasto-plastic models was introduced by Olmsted \cite{O08} who has added non-local terms terms to the Johnson-Segalman model \cite{JS77}. As in the phase field approach, the overall stress is decomposed into a fluid-borne stress, for instance Newtonian $\eta \dot \gamma$, and an elastic-borne stress $G \epsilon$, where $\eta$ is a viscosity and $G$ an elastic shear modulus. The equation governing the evolution of the deformation $\epsilon$ reads:
\begin{equation}
T \dot \epsilon=T \dot \gamma -{\mathcal I}(\epsilon)+\ell^2 \vec \nabla^2 \epsilon,
\label{JS}
\end{equation}
where $T \sim \eta/G$ is the relaxation time-scale of the components (e.g. the polymer). The diffusion term $\ell^2 \vec \nabla^2 \epsilon$ is responsible for non-locality. The function ${\mathcal I}$ reflects the elastic properties of the components and determines the constitutive relation measured in a homogeneous steady state. For the Herschel-Bulkley relationship (\ref{HB}) (we recall that $I=T\dot \gamma$), one obtains:
\begin{equation}
{\mathcal I}(\epsilon)=\left[\frac1a\;{\rm max}\left(0,\frac{G\epsilon}{\sigma_y}-1\right)\right]^{(1/n)}.
\end{equation}
The standard Johnson-Segalman model corresponds to the exponent $n=1$. The ratio $\epsilon_y=\sigma_y/G$ is the yield strain above which plastic events nucleate. The same model has successfully been used to describe complex fluids subject to shear-banding, with a vanishing yield stress \cite{O08,ORL00}. 

In the steady state, the elastic deformation obeys the non-linear Laplace equation $\ell^2 \vec \nabla^2 \epsilon -{\mathcal I}(\epsilon)=-T \dot \gamma$, which leads to exponential relaxations in space (see section~\ref{differences}). In the limit where $T$ becomes much smaller than $\dot \gamma^{-1}$, the same equation holds at all times and the dynamics becomes truly non-local in the sense (i).

The above relation was modified by Marmottant and Graner to model dry foams \cite{MG07}, assuming that the elastic part does not evolve over to the internal time $T$ but over the time-scale $\dot \gamma^{-1}$, which is assumed to provide the only time-scale of the problem. This assumption is close to that made by Forterre and Pouliquen \cite{PF09} for granular flows. Keeping the structure of Eq.~(\ref{JS}), the governing equation then takes the form:
\begin{equation}
\dot \epsilon=\dot \gamma \left(1-{\mathcal I}(\epsilon)+\ell^2 \vec \nabla^2 \epsilon\right),
\end{equation}
where ${\mathcal I}$ is a function which is essentially $0$ at low $\epsilon$ and which sharply increases and crosses $1$ at the yield strain $\sigma_y/G$. Note that the derivation and tests proposed in \cite{MG07} actually do not take the non-local term into account (they have $\ell=0$). We introduce it here, to clarify the connections between the various models.

\subsection{Fluidity as a Maxwell relaxation rate}
Considering again the ideal linear Couette cell controlled at imposed shear stress $\sigma$, the mechanical response of a soft material usually presents a transient in time. Over time-scales comparable to $\dot \gamma^{-1}$ or smaller, the structure does not have time to evolve: the time dependence is said to reflect \emph{visco-elasticity}. If the rheology evolves over time-scales long compared to $\dot \gamma^{-1}$, the system is said \emph{thixotropic}. Thixotropy is very close for time dependence to non-locality for space dependence. It reflects ageing, a property often shared by many soft glassy materials arrested in the glassy regime. Starting from a visco-elastic rheology relation, Derec et al. have proposed a model suited to describe such a thixotropic behaviour \cite{DAL01}. The fluidity $f$ is then introduced as a relaxation rate in a standard Maxwell visco-elastic model :
\begin{equation}
\dot \epsilon=-f \epsilon+ \dot \gamma.
\label{Derec}
\end{equation}
The auxiliary equation that governs the evolution of $f$ is macroscopic and implies ageing and shear rejuvenation processes. By contrast, Olmsted model assumes a constant time for the relaxation and Marmottant and Graner a time-scale inversely proportional to $\dot \gamma$. The three models thus differ by the identification of the relevant strain relaxation time. Interestingly, in the solid phase (i.e. below the Coulomb threshold) sheared granular packing display ageing that can be described by this equation and the fluidity parameter $f$ is directly related to the rate of localised plastic events, called \emph{hot-spots} that were directly visualised \cite{ANBCC12}. We will turn back to this point later.

\subsection{Fluidity as the rate of plastic events}
Close enough to the yield stress, flow occurs in concentrated emulsions and foams via a succession of reversible elastic deformations \cite{CDW15,DW15} and avalanches of irreversible plastic rearrangements (also called shear transformation zones in the literature \cite{FL98}). Such localised plastic events induce a long range anisotropic relaxation of the elastic stress over the system, which constitutes an obvious source of non-locality. Based on this observation, the fluidity in a class of models has been associated to the rate of plastic events, coarse grained in space and time. The physical picture assumes that the material is essentially in an elastic state under stress but due to disorder or temperature or mechanical fluctuations, localised plastic events nucleate, which induce local irreversible stress relaxation processes. The resulting deformation is absorbed elastically by the medium. This is at the origin of a global plastic deformation rate. For a given fast and local relaxation process, other plastic events may be triggered. When there is a continuous rate of coupled plastic events spanning a significant time, this is called an avalanche \cite{F98,TLB06,HKLP10,DBZU11}. This avalanching process occurs preferentially at larger stress. The exact account for avalanche dynamics as a net contribution to the final plastic deformation rate is a difficult issue and has therefore been, in most models, phenomenologically modelled by means of non-linear terms entering the fluidity equation.

The auxiliary equation for $f$ can be a stochastic Smoluchowsky equation \cite{HL98} but macroscopic --~essentially mean-field~-- models have also been derived explicitly. For example Bocquet et al. \cite{BCA09} have proposed a model, called KEP (kinetic elasto-plastic), which adds a spatial coupling to the probabilistic framework introduced by H\'ebraud and Lequeux \cite{HL98}. The model describes the evolution of the probability to observe locally a certain local "shear stress" and furthermore assumes that plastic events are triggered above a non-fluctuating "local yield stress". The macroscopic rheology turns out to be controlled by the behaviour of the shear stress probability distribution function in the immediate vicinity of the local yield stress: in a steady homogeneous state, a Herschel-Bulkley rheology is recovered.

In the KEP model, plastic events lead to a noise around them that helps to trigger other plastic events. Non-locality therefore appears as a dependence of a "local stress diffusion" on the rate of plastic events around the region considered. Both the elastic strain (and therefore the stress) and the rate of plastic events stem from an integral over the same probability distribution function. In the steady state (but not during transients), they can be related to each other. One recovers the relation between the rate $f$ of plastic events and elastic strain of Eq.~(\ref{Derec}), for $\dot \epsilon=0$:
\begin{equation}
f=\frac{\dot \gamma}{\epsilon}.
\label{reffeps}
\end{equation}
The non-local equation has been derived by Bocquet et al. in the steady state (and only in this case) and takes the form of a non-linear Poisson equation, analogous to that of Olmsted's model (\ref{JS}) in the steady state:
\begin{equation}
0= \dot \gamma -T^{-1}{\mathcal I}(f)+\ell^2 \vec \nabla^2 f.
\label{JSGamma}
\end{equation}
We later refer to this equation as the KEP constitutive relation or KEP model. However, a mathematical analogy is not a physical equivalence: the fluidity defined as the variable that appears in the non-local Laplacian term is, in one case, the elastic deformation $\epsilon$ and in the other, its relaxation rate $f$. Furthermore, to fully solve the problem in association with a Laplacian term in the formulation, it is necessary to provide boundary conditions for fluidity \cite{MBC14}. The choice of the expression for the fluidity must then be consistent with the physical boundary conditions. We will turn later on the possible tests to determine which fluidity is the relevant one.

Importantly, the macroscopic emergence of the auxiliary fluidity equation involves a Laplacian operator which physically represents the spatial range of plastic relaxations. Mathematically, it is the source of non-locality in the constitutive relation.
A possible issue is that the stress relaxation induced by plastic events can have in general an anisotropic character, even in a statistical sense, which is not reflected by the isotropic Laplacian term in (\ref{JSGamma}). If, in this triggering process, anisotropy is important, higher order terms must be included in the spatial expansion of the stress propagator in \cite{BCA09}.
Also, close to the yield condition, long range avalanches may take place and spatial coupling can span large distances that eventually diverge at the yield point. In this limit, the Laplacian, which is essentially a mean field operator, is unlikely to capture non-locality, as it is well known that Landau-like approaches generically fail in the vicinity of a critical point.

\subsection{Fluidity as the inverse of viscosity}
Bocquet et al. \cite{BCA09} have made a further, apparently innocent, step: as the elastic stress $\sigma$ is proportional to the elastic strain $\epsilon$, the structure of the equations does not change if $f$ is defined as
\begin{equation}
f=\frac{\dot \gamma}{\sigma}
\end{equation}
instead of Eq.~(\ref{reffeps}). The fluidity $f$ becomes the inverse of the particle-borne viscosity. The equation governing $f$ remains of the form (\ref{JSGamma}). We will discuss below the problems associated with the change of variable $\epsilon \to \sigma$. We emphasise again that the relation between the rate of plastic events and the fluidity is not trivial and \emph{must be tested}. 

To handle granular matter, Kamrin et al. \cite{KK12} have proposed to rescale this inverse viscosity by the yield stress $\sigma_y$. If $\sigma_y$ is constant, this does not change the shape of the equation governing the new fluidity (Eq.~\ref{JSGamma}):
\begin{equation}
f=\frac{\sigma_y}{\sigma}\,\dot \gamma.
\end{equation}
The fluidity $f$ is then homogeneous to a strain rate. Again, the function ${\mathcal I}$ can be determined from the constitutive relation measured in a homogeneous steady state. For the Herschel-Bulkley relationship, one obtains the implicit equation:
\begin{equation}
\frac{{\mathcal I}(f)}{1+a {\mathcal I}^n(f )}=T f,
\label{ImplicitEqformathcalI}
\end{equation}
whose solution takes the form:
\begin{equation}
{\mathcal I}(f)= T f+a (T f)^{1+n}+{\mathcal O}(f^{1+2n}).
\end{equation}
For the particular case of frictional granular material, for which $n=1$, one gets the analytical solution:
\begin{equation}
{\mathcal I}(f)= \frac{T f}{1-a T f} \, .
\label{explicitnone}
\end{equation}
%

\section{What are the differences between non-local constitutive relations proposed for granular matter?}
\label{differences}

\subsection{Fluidity, from soft-matter to granular matter}
As seen above, the fluidity in Kamrin et al.'s model for granular matter \cite{KK12}, derived from the results of Bocquet et al. \cite{BCA09}, writes
\begin{equation}
f=\frac{\sigma_y}{\sigma}\,\dot \gamma=\frac{\dot \gamma}{\mathcal Y}=\frac{\mu_c P}{\sigma}\,\dot \gamma.
\label{fKamrin}
\end{equation}
The auxiliary equation for this parameter is derived from Eq.~(\ref{JSGamma}) by a linearisation that we discuss below in details, as it is problematic.

The transposition from soft matter to rigid grains poses a fundamental issue. The fluidity must obviously be a state variable. By \emph{state variable}, we mean that the fluidity $f$ must be a coarse-grained field (in space and time) which can be determined from the state of the system. It can for example depend on the strain, the strain rate, the volume fraction, the mean number of contacts. For a granular material, it can also involve the fraction of sliding contacts or the orientation of the contacts. For foams or emulsions, it can reflect the elastic deformation of elementary cells. However, it cannot depend explicitly on the stress tensor, which is not a state variable itself. This directly results from Newton second's law, which tells that positions and velocities of the particles determines the state of a mechanical system, from which forces are derived. Similarly, the (non-local) constitutive relation must relate the stress tensor to the state variables --~and not the opposite. $f$ as defined in Eq.~(\ref{fKamrin}) was a state variable for an elastic system, because it was fundamentally based on the elastic deformation. It is no longer the case for a granular system composed of rigid grains.

\subsection{Does flowing granular matter exhibit plastic events?}
The transposition of models derived for elasto-plastic material to granular systems obviously requires that granular matter behaves as hypothesised by elasto-plasticity. The elasto-plastic picture assumes that the material behaves most of the time like a solid, but presents local and short-lived plastic events \cite{FL98,TLB06}. The associated scenario is a localised rupture initiation followed by a scale free avalanche of localised events. In order to investigate whether this picture constitutes an alternative to the jamming scenario to interpret the non-local nature of the granular rheology, we compare, by means of numerical simulations (Discrete Element Method), the dynamics of two otherwise identical systems composed of hard and soft grains.

The general numerical set-up is that used in \cite{BTCCA13}: we consider a two-dimensional system composed of $\sim 2.10^3$ spherical particles of a mean diameter $d$, with a $\pm 20\%$ polydispersity. Such a choice ensures that the sample will not crystalise, The particles can interact through contact forces modelled as a viscoelastic force along the normal contact direction and as a Coulomb friction along the tangential direction. The corresponding coefficient of restitution is $e \simeq 0.9$. The Coulomb friction coefficient is set to $\mu_p=0.4$ for fictional particles and $\mu_p=0$ for the frictionless system. The particles are confined in a plane shear cell composed of two rough solid walls made by the same particles, glued together. Periodic boundary conditions are used along the shear direction $x$. The position of the wall is controlled in order to impose a constant normal stress $P$ and constant and opposite velocities of the walls along $x$. The system is in the asymptotic rigid limit when the ratio $k_n/P$ of the normal spring constant with the pressure is sufficiently large (typically above $10^3$).

The presence of localised plastic events is usually based on a visual inspection of different fields. The squared deviation from an affine deformation on a local scale has for instance been proposed as a field indicating plastic activity \cite{FL98,LC09}. However, such a quantity, as well as all those based on the non-affine velocity, characterises fluctuations around the mean flow, and not the local contribution of a certain area to the mean flow. We wish here to propose a practical definition of these events, based on the quantitative criterion that they must be separated in time and localised in space. Importantly, to match their role played in elasto-plastic models \cite{FL98,LC09,BCA09,HK13}, they must also contribute additively to the average shear rate $\dot \gamma$.

\begin{figure}[t!]
\includegraphics[scale=0.25]{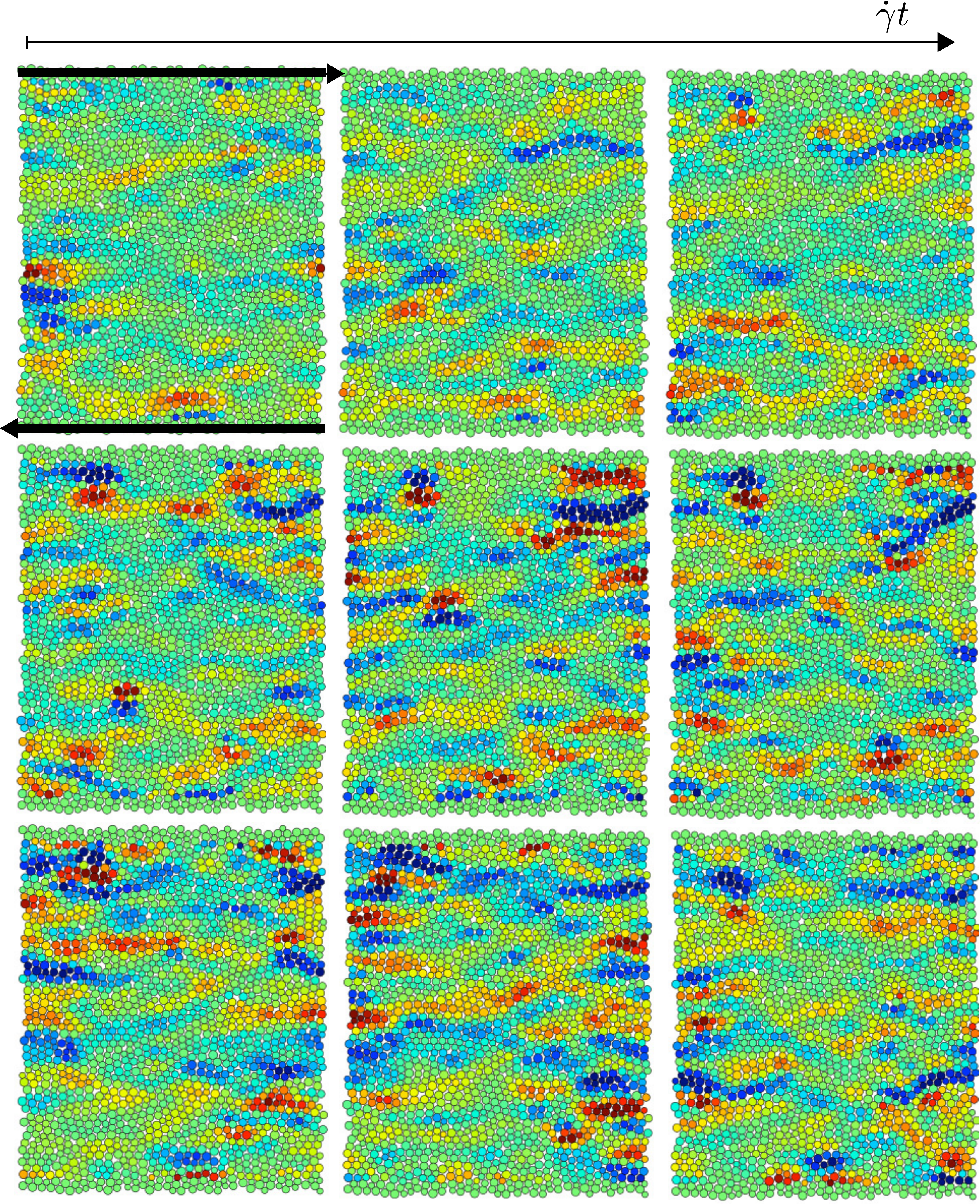}
\caption{(Color online) Image sequence showing the temporal evolution of the local shear rate $\dot \Gamma$ for a system of frictional hard grains ($kn/P=2 \, 10^4$) in the quasi-static limit ($I=5 \,10^{-4}$). Similar results are found for frictionless particles. Color code from blue ($\dot\Gamma=-40\dot\gamma$) to red ($\dot\Gamma=40\dot\gamma$). Time lapse between two successive images: $\dot\gamma \Delta t = 10^{-3}$.}
\label{Fig4bis}
\end{figure}

\begin{figure}[t!]
\includegraphics[scale=0.25]{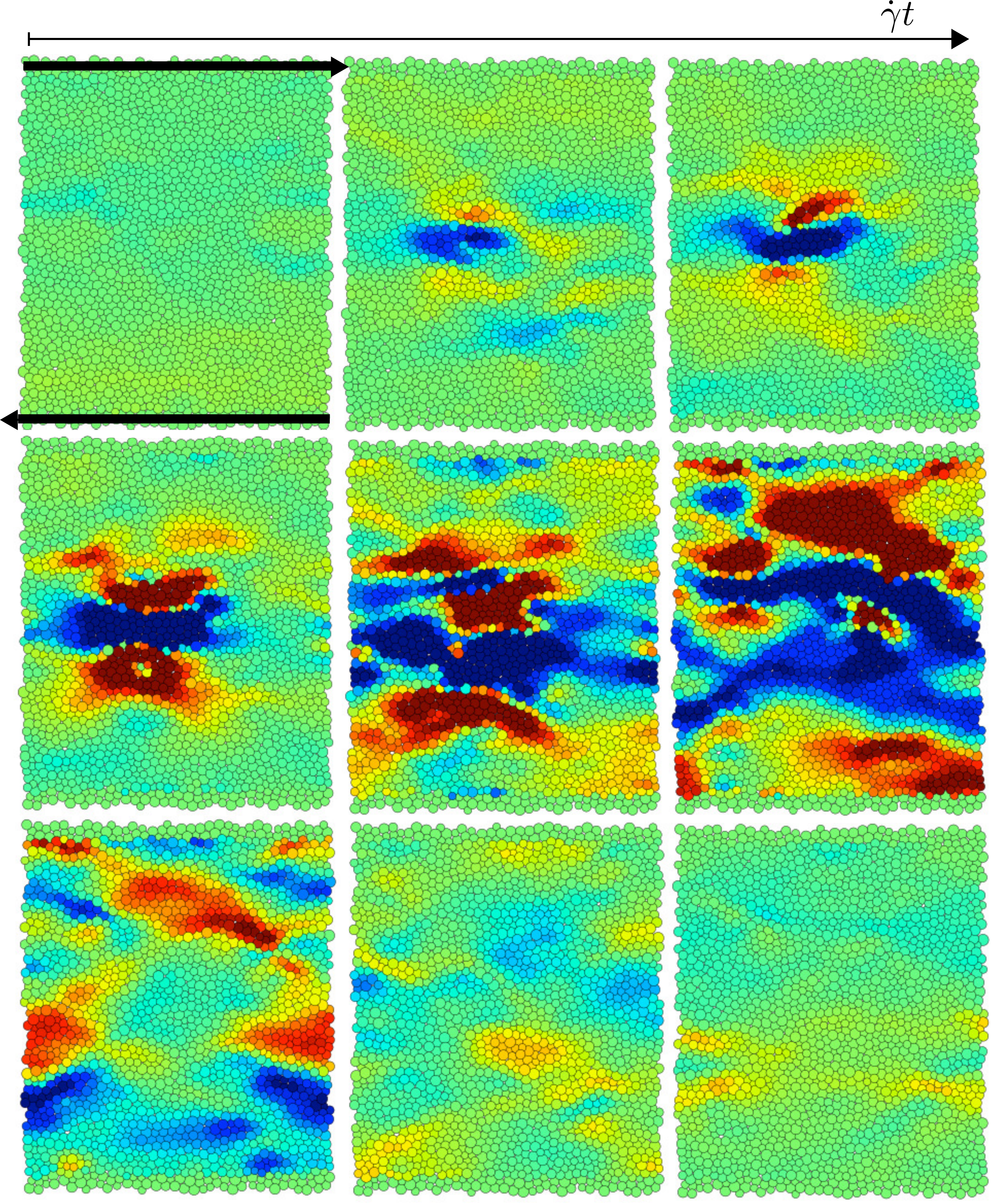}
\caption{(Color online) Image sequence showing the temporal evolution of the local shear rate $\dot \Gamma$ for a system of frictional soft grains ($kn/P=10$) in the quasi-static limit ($I=5 \,10^{-4}$). Similar results are found for frictionless particles. Color code from blue ($\dot\Gamma=-40\dot\gamma$) to red ($\dot\Gamma=40\dot\gamma$). Time lapse between two successive images: $\dot\gamma \Delta t = 10^{-3}$.}
\label{Fig5bis}
\end{figure}

\begin{figure}[t!]
\includegraphics{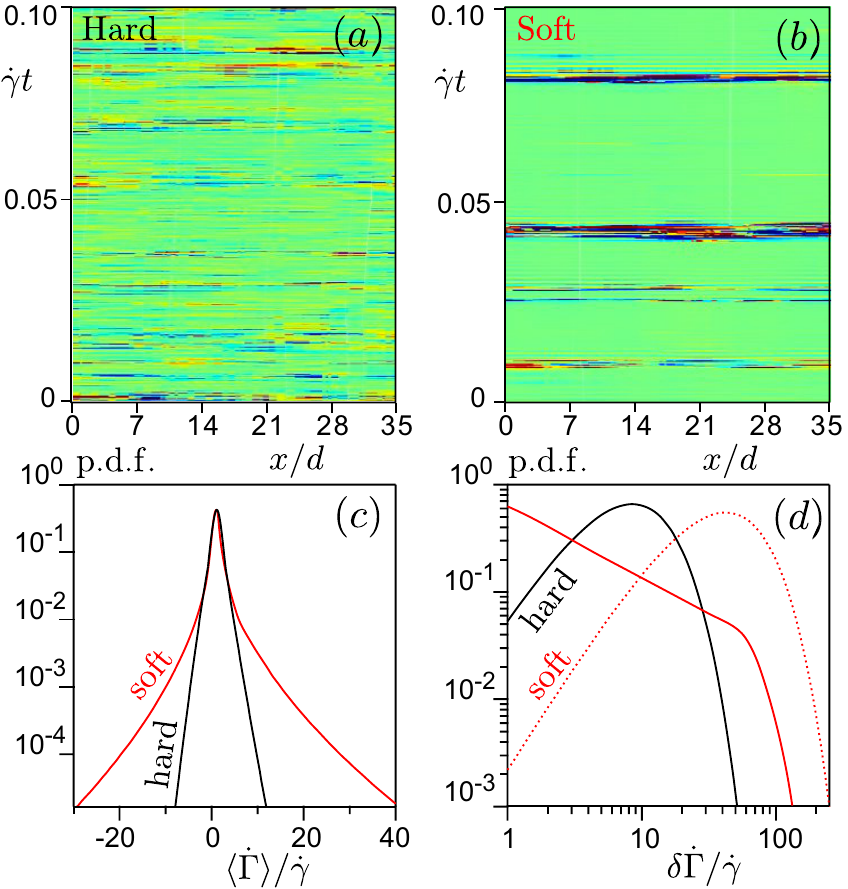}
\caption{(Color online) Space-time diagrams showing the local contribution of $\dot\Gamma$ to the shear rate $\dot \gamma$, measured on the central line of the cell, for a system of sheared hard (a) and soft (b) grains. For this example, the ratio of the grain contact stiffness $k_n$ to the overall pressure $P$ is $2 \, 10^4$ and $10$ respectively; the shear rate corresponds to an inertial number $I=5 \, 10^{-4}$. Color code from blue ($\dot\Gamma=-40\dot\gamma$) to red ($\dot\Gamma=40\dot\gamma$). (c) Probability distribution function (PDF) over time of the space average $\langle\dot \Gamma \rangle$, for hard (black line) and soft (red line) grains. (d) PDF of the spatial standard deviation $\delta \dot\Gamma$. Red dotted line: $\delta \dot \Gamma$ computed when $|\langle \dot \Gamma \rangle|\geq 5 \dot \gamma$ (soft system).}
\label{Fig3}
\end{figure}

In order to detect localised plastic events, we have built a coarse-grained field $\dot \Gamma(\vec r,t)$ reflecting, at time $t$, the local contribution to $\dot \gamma$ of a small region around the position $\vec r$. We impose that the time average of $\dot \Gamma$ must everywhere give $\dot \gamma$. A coarse-graining method similar to that proposed for the stress tensor \cite{GG01,GG02} is adapted here to the computation of velocity differences and we take:
\begin{eqnarray}
\dot\Gamma(\vec{r},t)=\frac{\sum \limits_{j=1}^N[u_i(\vec{r},t)-u_j(\vec{r},t)][z_i(t)-z_j(t)]\exp(\frac{-||\Delta\vec{r}||^2}{2\delta^2})}{\sum \limits_{j=1}^N[z_i(t)-z_j(t)]^2\exp(\frac{-||\Delta\vec{r}||^2}{2\delta^2})},
\label{BigGammaDotDef}
\end{eqnarray}
where $u_i(\vec{r},t)$ is the velocity of the grain $i$ at the time $t$ and $||\Delta\vec{r}||=\sqrt{[z_i(t)-z_j(t)]^2+[x_i(t)-x_j(t)]^2}$ denote the distance between the grain $i$ and $j$. $\delta$ is the coarse-graining length, typically on the order of the grain size $d$. We display in Fig.~\ref{Fig4bis} and \ref{Fig5bis}, for a system of rigid and soft grains respectively, the map of the local contribution $\dot \Gamma$ to the shear rate $\dot \gamma$ at different times. In Fig.~\ref{Fig3}a,b we show corresponding spatio-temporal diagrams built on the central line of the cell. We observe contrasted behaviours in the two cases. In the soft system, nothing much happens most of the time, except for short periods of intense activity, associated with a cascade of plastic events. Conversely, the hard system presents more moderate but permanent fluctuations even for asymptotically small $\dot \gamma$.

To make these observations quantitative, Fig.~\ref{Fig3}c shows the probability distribution function (PDF) over time of $\langle \dot \Gamma\rangle$, which is the space average of $\dot \Gamma$ over the cell. In panel (d), we similarly display the PDF of the spatial standard deviation $\delta \dot \Gamma$. The hard-particle system presents a narrow Gaussian distribution of $\langle \dot \Gamma\rangle$ around $\dot \gamma$, while the PDF corresponding to the soft system shows stretched tails, which are due to a very intermittent behaviour associated with these plastic events. The PDF of $\delta \dot \Gamma$ provides informations about spatial heterogeneities in the system. The peak of the black line around $10 \dot \gamma$ in Fig.~\ref{Fig3}d indicates that they are large and permanent in the hard system. For the soft system, the PDF shows an algebraic decay, which means that the field $\dot \Gamma$ is homogeneous most of the time. However, when the computation of $\delta \dot \Gamma$ is restricted to the periods of time where plastic events occur (periods where $|\langle \dot \Gamma \rangle|$ is larger than a given value, here $5 \dot \gamma$ in Fig.~\ref{Fig3}d), its PDF also presents a peak: in the soft system, plastic events are associated with a very heterogeneous field of $\dot \Gamma$. Conversely, an assembly of rigid particles does not present local plastic events when sheared permanently. Its dynamics is not intermittent but presents spatial heterogeneities.

In conclusion, when constituted of rigid particles, sheared granular systems do not present a succession of elastic energy accumulation and sudden release. Their dynamics rather show permanent cooperative motions. As a consequence, approaches explicitly based on elasto-plasticity developed for soft systems, such as those discussed above, cannot be transposed to granular flows, where elasticity of the grains is irrelevant. The physical foundation Kamrin et al.'s approach for granular matter \cite{KK12} seems already in this perspective, extremely problematic.
 
\subsection{Linearisation in the case of a homogeneous shear stress profile}
The simplest situation in which the model predictions can be tested is a shear cell inside which the shear stress, and therefore the yield parameter $\mathcal{Y}$, are homogeneous. In such a situation, a local rheology predicts a constant shear rate, which, once rescaled by $T$, is denoted $I_\infty$. This cell is driven by boundary layers on each side, whose properties are not necessarily those of the bulk. How to realise this in practice for numerical simulations is for example described in Bouzid et al. \cite{BTCCA13}.

We first consider the case where $I_\infty$ does not vanish. By definition, we have $\mathcal{Y}=\mu(I_\infty)$. Making profit that $\mathcal{Y}$ is constant, one can linearise the equations in $I$ around $I_\infty$. Using the gradient expansion model (Eq.~\ref{Mehdi}), where $I$ is the fluidity, we obtain:
\begin{eqnarray}
\ell^2 \vec \nabla^2 (I-I_\infty)&=&n \left((1+a I_\infty^n)^{-1}-1\right)(I-I_\infty)\nonumber\\
&=&n \frac{{\mathcal Y}-1}{{\mathcal Y}}(I-I_\infty).
\end{eqnarray}
Denoting by $z$ the axis transverse to the flow, we obtain exponential solutions of the form
\begin{eqnarray}
I=I_\infty +A_+ \exp(z/L)+A_- \exp(-z/L),
\end{eqnarray}
where the relaxation length $L$ is given by:
\begin{equation}
L^2=\frac{1+anI_\infty^n}{anI_\infty^n}\ell^2=\frac{\ell^2 {\mathcal Y}}{n({\mathcal Y}-1)}\quad{\rm for}\quad \mathcal{Y}>1.
\label{L2MehdiYsup1}
\end{equation}
It is important to emphasise the status of this length $L$. As ${\mathcal Y}$ is the control parameter of this particular thought experiment (or numerical simulation \cite{BTCCA13}), $L$ can be expressed as a function of ${\mathcal Y}$. It does not make ${\mathcal Y}$ a state variable which would control another state variable $L$. Note also that $L$ can equally be expressed as a function of $I_\infty$.

Consider now the KEP equation~(\ref{JSGamma}) with the fluidity $f$ proposed by Kamrin et al. (\ref{fKamrin}). As above, making profit that ${\mathcal Y}$ is homogeneous, this equation can be linearised around $f_\infty$ as:
\begin{equation}
\ell^2 \vec \nabla^2 (f-f_\infty)=\frac{\ell^2}{L^2}(f-f_\infty),
\end{equation}
where $L$ is now given by:
\begin{equation}
L^2=\ell^2\left(\frac{1}{\mathcal Y}+\frac 1{n ({\mathcal Y}-1)}\right)  \quad{\rm for}\quad \mathcal{Y}>1.
\label{Lupper}
\end{equation}
In the limit ${\mathcal Y} \to 1$, the relaxation length takes exactly the same form $L \sim \ell/\sqrt{n ({\mathcal Y}-1)}$ for the two models, despite their differences.

The second case corresponds to ${\mathcal Y}<1$, so that $I_\infty=0$. Again, the equations can be expanded around $I=I_\infty$ but this linearization is completely different from the previous one, as $I=0$ is not solution of the equations: one needs the non-local term to get a solution. With the KEP model (\ref{JSGamma}), one obtains:
\begin{equation}
\ell^2 \vec \nabla^2 f=(1-{\mathcal Y})f,
\end{equation}
which gives exponential relaxations over a length $L$ given by
\begin{equation}
L^2= \frac{\ell^2}{(1-{\mathcal Y})} \quad{\rm for}\quad \mathcal{Y}<1.
\label{Llower}
\end{equation}
Using the gradient expansion model (\ref{Mehdi}), we get:
\begin{equation}
\ell^2 \vec \nabla^2 I=\chi^{-1}({\mathcal Y})I.
\end{equation}
The shear rate therefore relaxes over a length $L$ given by:
\begin{equation}
L^2=\frac{\ell^2}{\chi^{-1}(\mathcal{Y})} \quad{\rm for}\quad \mathcal{Y}<1.
\label{ellbelow}
\end{equation}
In the vicinity of the critical conditions, $\kappa$ is indeed small, so that the linear approximation $\chi(\kappa)\simeq 1- \kappa$ can be used. The divergence of $L$ at $\mathcal{Y}\to 1$ is therefore given, again, by eq.~(\ref{Llower}). 

\begin{figure*}[t!]
\includegraphics{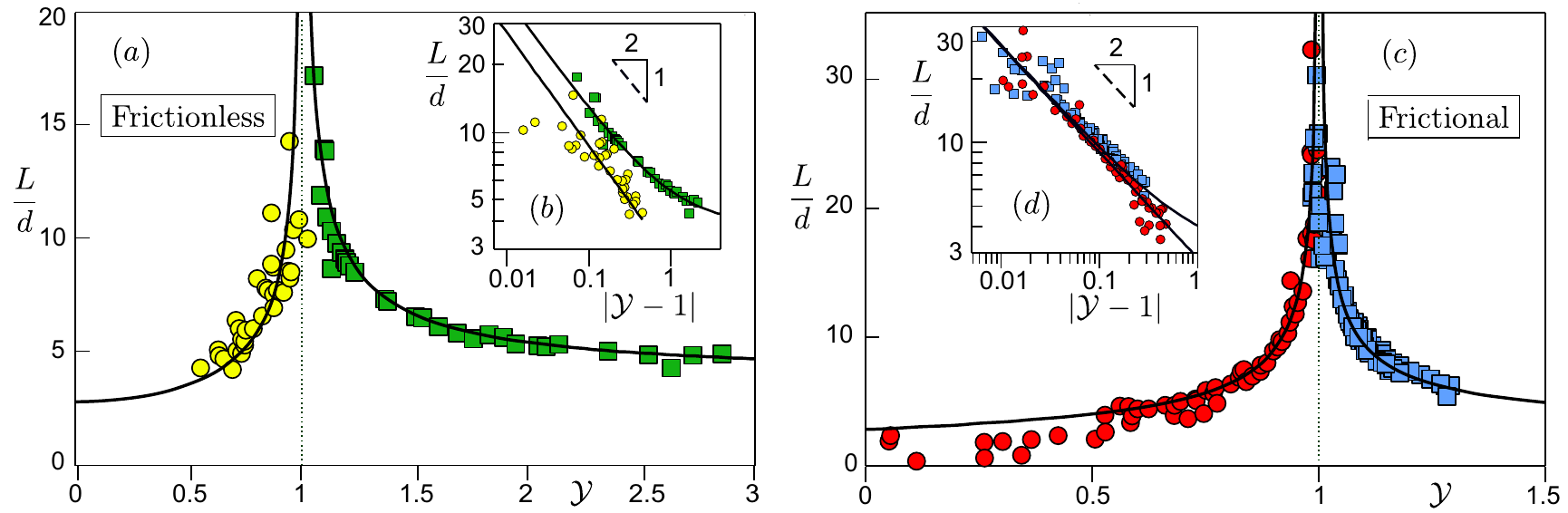}
\vspace{- 0 mm}
\caption{Relaxation length $L$ for frictionless (a) and frictional grains (c) below (red and yellow circles) and above (blue and green squares) yield conditions, data from \cite{BTCCA13}. Solid lines: fit of the data to Eqs.~(\ref{L2MehdiYsup1}) and (\ref{ellbelow}), diverging as $|\mathcal{Y}-1|^{-1/2}$ when $\mathcal{Y} \to 1$. (b) and (d) Log-log plot of the same quantities.}
\vspace{0 mm}
\label{Fig4}
\end{figure*}

\begin{figure}[t!]
\includegraphics[scale=0.8]{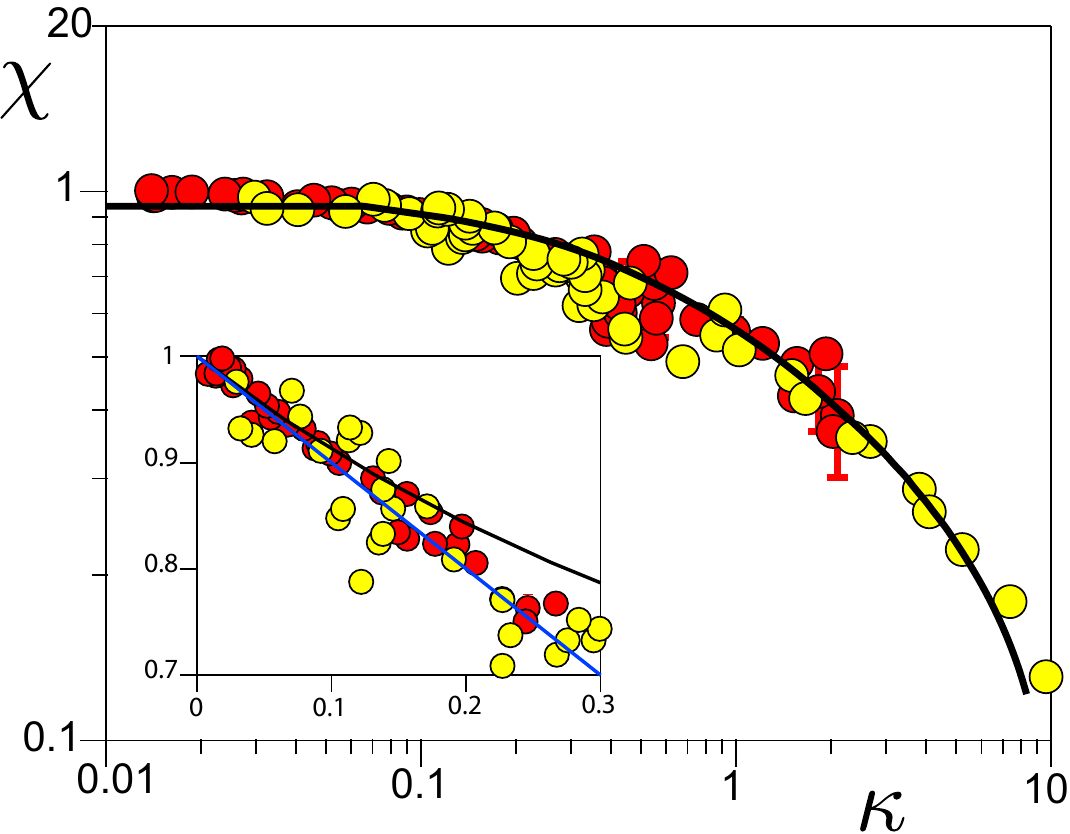}
\vspace{0 mm}
\caption{Function $\chi(\kappa)$ numerically measured for $\mathcal{Y}<1$ for frictional (red circles) and frictionless (yellow circles) grains -- the system is that described \cite{BTCCA13}. The black solid line is a fit with the empirical expression $ \chi=\frac{\sqrt{(1-\kappa \alpha)^2+\kappa \beta (\kappa \alpha-1)}}{1 - \alpha \kappa}$ with $\alpha=-15.95$ and $\beta=16.30$. Inset: zoom on the small values of $\kappa$, in Lin-Lin axes.}
\vspace{0 mm}
\label{Fig5}
\end{figure}

As an illustration, we display in Fig.~\ref{Fig4} such diverging relaxation lengths extracted from numerical simulations of sheared layer \cite{BTCCA13}. Velocity data, such a presented in Fig.\ref{Fig1}b can be obtained systematically for $\mathcal{Y}$ values above and below $\mathcal{Y}=1$. The fit of the velocity profiles is made with a function of the form $\dot\gamma_b z + C \sinh(z/L)$,  where $C$ and $L$ are adjustable. This provides a direct measurement of a relaxation length $L$, which effectively diverges on both sides of the critical point $\mathcal{Y}=1$ according to the theoretical predictions (\ref{L2MehdiYsup1}) and (\ref{ellbelow}). Fig.~\ref{Fig5} shows the shape of the function $\chi(\kappa)$. Its non-linear behaviour, which roughly starts when $\kappa > 0.1$, is at the origin of the asymmetry of $L$ with respect to the yield point $|\mathcal{Y}|=1$ when sufficiently far away from this point.

In conclusion, as far as linearisation is concerned, the two models lead, despite their distinct starting point, exactly to the same predictions for constant stress conditions in the vicinity of the yield condition $\mathcal{Y}=1$. Because two equations giving the same exponential solutions are not necessarily equivalent, this shear cell configuration in which the yield parameter $\mathcal{Y}$ is controlled and homogeneous thus cannot be used to discriminate between the different possibilities to build a fluidity. Further tests focusing on  time transients and on heterogeneous situations are needed to test the starting constitutive equations, which do not reduce to their linearised expressions.

\subsection{Does the KEP rheology reflect a dynamical phase transition?}
An important claim made by Bocquet et al. in \cite{BCA09} is that the KEP constitutive relation reflects a dynamic phase transition controlled by the stress, in relation to the divergence of the relaxation length $L$ on both sides of ${\mathcal Y}=1$. Because the model used by Kamrin et al. \cite{KK12} is derived from the KEP approach, this claim would also apply to granular matter. On the opposite, in our framework, we argue that the non-local rheology describes the same liquid phase above and below the yield conditions. Because this controversy concerns an essential point of physics, we find it important to develop in this subsection some technical but essential details of this issue.

As already mentioned, the starting KEP equation (\ref{JSGamma}), when linearised around a homogeneous stress state corresponding to a given constant $\mathcal{Y}$, leads to the generic equation:
\begin{equation}
\vec \nabla^2 f=\frac{f-f_\infty}{L^2} \, .
\label{GLEq}
\end{equation}
$f_\infty$ and $L$ depend on the value of $\mathcal{Y}$. This is the Ginsburg-Landau equation used by Kamrin et al. \cite{KK12} with the fluidity given by Eq.~(\ref{fKamrin}). A crucial point is that this equation is used by these authors in \emph{non-homogeneous} situations, i.e. taking for $\mathcal{Y}$ the \emph{local} value, with functions $f_\infty({\mathcal Y})$ and $L({\mathcal Y})$ \cite{HK13,HK14}. This would be a correct assumption for a slowly varying stress field if $\mathcal{Y}$ was a state variable. We have discussed above why it is not the case, even though it is the control parameter of the considered linearisation. 

Let us further illustrate the mathematical differences between two seemingly equivalent derivations. Consider the KEP constitutive relation (\ref{JSGamma}) with $\dot\gamma = f \mathcal{Y}$ (Eq.~\ref{fKamrin}). Associated with (\ref{ImplicitEqformathcalI}), which determines $\mathcal{I}(f)$ in the homogeneous case, and whose solution is given by (\ref{explicitnone}) for $n=1$, we obtain:
\begin{equation}
\ell^2 \vec \nabla^2 f=f\left[\frac{1}{1-a T f}-{\mathcal Y}\right].
\label{case1}
\end{equation}
However, let us alternatively start from (\ref{GLEq}) and plug in expressions of $f_\infty$ and $L$. $f_\infty$ is the solution of the equation $\dot\gamma = T^{-1} \mathcal{I}(f_\infty)$, i.e. $f_\infty({\mathcal Y})=({\mathcal Y}-1)/(a T {\mathcal Y})$.  The expression for $L$ depends on whether $\mathcal{Y}$ is larger (Eq.~\ref{Lupper}) or smaller (Eq.~\ref{Llower}) than unity. To make it compact, let us focus close to the yield condition $\mathcal{Y}=1$ for which both cases can be summed up with $L({\mathcal Y})^2 \sim \ell^2/|{\mathcal Y}-1|$. Doing so, instead of (\ref{case1}), we obtain:
\begin{equation}
\ell^2 \vec \nabla^2 f=|{\mathcal Y}-1| \left [f-\frac{{\mathcal Y}-1}{a T {\mathcal Y}} \right ] .
\label{case2}
\end{equation}
Equations (\ref{case1}) and (\ref{case2}) are obviously not the same. In particular, their behaviour is very different  when $\mathcal{Y} \to 1$ as the right hand side vanishes in the second case but stays finite in the first one. In other words, the original KEP equation (\ref{JSGamma}) and the final Ginsburg-Landau equation used by Kamrin et al., are irreducible one to the other: the transformation is neither mathematically nor physically justified.

The divergence of the relaxation length $L$ on both sides ${\mathcal Y}=1$ has been interpreted as the signature of a dynamic phase transition controlled by ${\mathcal Y}$. The region ${\mathcal Y}<1$ would correspond to the solid-like behaviour while the region ${\mathcal Y}>1$ would correspond to the liquid-like behaviour. However, the original KEP model (\ref{case1}) was entirely based on the description of a single liquid-like phase and the Ginsburg-Landau equation (\ref{case2}) used by Kamrin et al. \cite{KK12} is not a controlled approximation of it. ${\mathcal Y}$, which is not a state variable, cannot be the parameter controlling the phase transition. The liquid flowing phase can exist even below the yielding conditions, for ${\mathcal Y}<1$.

\subsection{Fluidity and boundary conditions}
\begin{figure}[t!]
\includegraphics[scale=0.6]{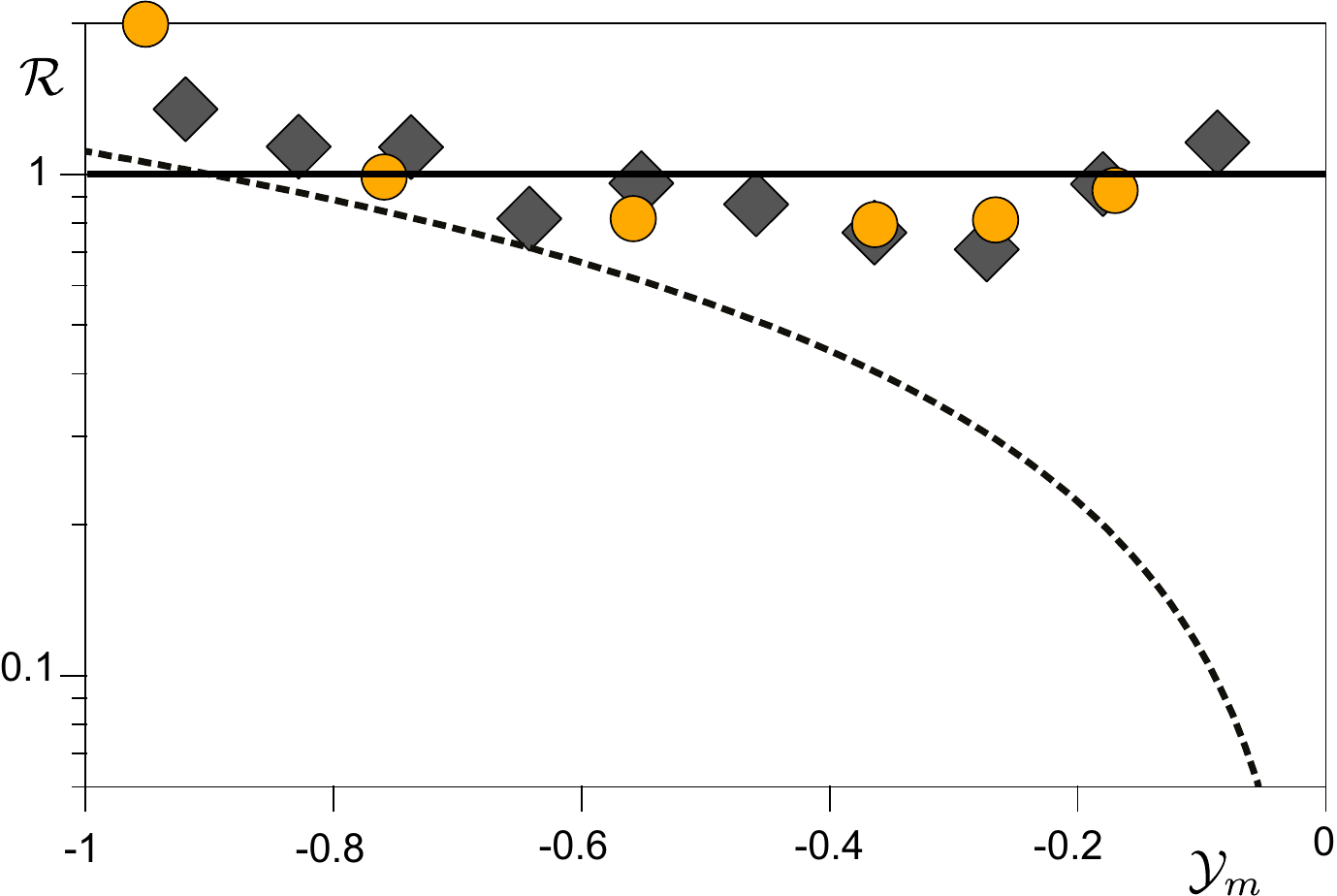}
\vspace{0 mm}
\caption{Ratio $\mathcal{R}$ of the absolute value of the shear rate on the outer and inner sides of a stress discontinuity, as a function of the yield parameter $\mathcal{Y}_m$. Yellow circles are for frictionless grains and gray lozenges are for frictional grains. Thick solid line $\mathcal{R}=1$: prediction of the gradient expansion model \cite{BTCCA13} using $I$ as fluidity. Dotted line: prediction of fluidity theory \cite{KK12} using $f=\dot{\gamma}/\sigma$ as a fluidity parameter.}
\vspace{0 mm}
\label{Fig6}
\end{figure}

An important consequence of the choice of a particular fluidity $f$ is the underlying assumption that $f$ and its gradient $\vec \nabla f$ are continuous, otherwise the use of the Laplacian operator would not make any sense. This remark provides a constraint on the nature of $f$, which can for example be tested in a situation where the stress varies extremely rapidly in space between two states. In this spirit, we have performed numerical simulations where a secondary micro-rheometer is placed in the bulk of a shear cell as presented previously (see Figs.\ref{Fig1}b and \ref{Fig2} and details in Bouzid et al. \cite{BTCCA15}). Shearing within the micro-rheometer is obtained by means of localised bulk forces along two lines which induce a discontinuity of the shear stress. We have measured numerically the ratio $\mathcal{R}$ of the absolute value of the shear rate on one side and on the other side of the stress discontinuity. The pressure remains constant and the direction of shearing is reversed ( $\dot \gamma$ changes sign at the discontinuity). Fig.~\ref{Fig6} shows that $|\dot \gamma|$ is indeed continuous ($\mathcal{R}=1$) in both frictionless and frictional cases. On the opposite, the fluidity proposed by Kamrin et al. (Eq.~\ref{fKamrin}) is not in agreement with the data, especially when the yield parameter approaches zero.

\section{Further tests: the question of dynamical mechanisms}
\label{furthertests}

We have presented a short review of non-locality in granular flows. We have mainly focussed on the comparison of two models: the KEP model adapted for granular matter  by Kamrin et al.\cite{KK12,HK13} (see Eq.~\ref{fKamrin}), and the gradient expansion that we have proposed \cite{BTCCA13} (see Eqs.~\ref{NLfriction} and \ref{Mehdi})). The difference between these models has not been recognised so far in the literature, essentially because they lead to the same predictions for the velocity profile in a situation where the stress is homogeneous. However, our conclusion is that these models are fundamentally different and that their differences can be tested.

These tests must be performed in situations that are strongly heterogeneous in space, like the one shown in Fig.~\ref{Fig6}, or to unsteady situations \cite{RTACS15}. They can of course concern the direct predictions of the model, but also the choice for the fluidity, which is not necessarily the inverse of viscosity, as well as the associated boundary conditions. Importantly, these conditions should not be fitted, but part of the physical analysis of the problem. More fundamentally, the difference between the models is to be found in the hypothesis made to derive them and in particular in the dynamical mechanisms underlying their dynamics. For instance, many different explanations have been proposed for the very same Herschel-Bulkley rheology. In the context of non-locality, let us give several example where ingredients could be tested. Different models assume the proportionality between the decay rate of fluidity and the rate of plastic events. Such a proportionality can therefore be investigated experimentally. Other models like KEP prescribe not only the average fluidity but also its distribution. The measurement of such a quantity is a more severe test than the fit of velocity profiles. Would a model assume the existence of microscopic yield conditions for the nucleation of plastic events, it would then be necessary to determine this quantity, to show that it exists and that it is constant.

In the case of granular matter, we have shown that the main point separating the KEP model and the gradient expansion model, is the existence or not of elasto-plastic localised events in the liquid regime. The KEP model is directly adapted from soft matter and assumes that elasticity dominates the dynamics. In the test presented here, we have numerically shown that, in the rigid limit, there are no localised plastic events and the flow is dominated by non-affine collective motion along soft modes. One could argue that the Coulomb friction condition at the contacts between the grains may lead to plastic events. However, comparing frictional and frictionless grains, we do not see any difference neither on non-locality nor on the absence of plastic events. Beyond other important reasons, we have shown that the fluidity proposed by Kamrin et al. \cite{KK12}, as an extension of the KEP model to rigid granular packings, is not a state variable and is thus not continuous across a stress discontinuity. We acknowledge that, in spite of the fact that our choice of the fluidity parameter for dense granular flows respects the state variable requirements and quantitatively predicts some situations, it does not clarify the understanding of the actual microscopic mechanisms at work to definitely unravel the question of non-local rheology. We recently discussed some limitations on such a choice and we pursue the work of identifying the microscopic or mesoscopic processes associated with the flow of hard grains \cite{BTCCA15,RTACS15}.  

Finally, amongst the points that have created a confusion in the literature, is the fact that granular matter does present localised plastic events, but \emph{only in the solid regime} \cite{ANBCC12,LBAMcNC14}, not in the liquid regime discussed here, were grain elasticity is irrelevant. Let us note that numerical simulations performed with the standard Coulomb model of friction at contacts, which perfectly reproduce observations in the dense liquid regime and in particular non-locality, are not able to reproduce creep in the solid regime. These two regimes (solid-like and liquid-like) must eventually be described, but the transition between these dynamical phases is known to be subcritical and to present a hysteresis, a key aspect of granular matter that remains unexplained at present. In this context, a non-local transition between solid and liquid was addressed by Wyart \cite{W09}, based on the generic outcome of the Maxwell rigidity transition for hard granular packing. Importantly, the KEP model claims to describe this dynamical phase transition as a critical transition controlled by the stress, and the rheology both above and below the transition. The gradient expansion, on the opposite, is based on the fact that shear stress cannot be a control parameter for this transition and describes the system as a unique continuous liquid phase both above and below yield conditions. This approach is therefore perfectly compatible with a subcritical transition to the solid regime, as it does not describe the later.


\begin{thebibliography}{}

\bibitem{AD01}  
B. Andreotti and S. Douady,
Phys. Rev. E  \textbf{63}, 031305 (2001).

\bibitem{R03}
J. Rajchenbach,
Phys. Rev. Lett. \textbf{90}, 144302 (2003).

\bibitem{J06}
J. T. Jenkins,
Phys. Fluids \textbf{18}, 103307 (2006).

\bibitem{A07}  
B. Andreotti,
Europhys. Lett. \textbf{79}, 34001 (2007).

\bibitem{GCOAB08}
J. Goyon, A. Colin, G. Ovarlez, A. Ajdari and L. Bocquet,
Nature. \textbf{454}, 84 (2008).

\bibitem{BCA09} 
L. Bocquet, A. Colin and A. Ajdari,
Phys. Rev. Lett. {\bf 103}, 036001 (2009).

\bibitem{GCB10}
J. Goyon, A. Colin and  L. Bocquet,
Soft Matter \textbf{6}, 2668 (2010).

\bibitem{CMCB12}
P. Chaudhuri, V. Mansard, A. Colin and L. Bocquet,
Phys. Rev. Lett. \textbf{109}, 036001 (2012).

\bibitem{MC12}
V. Mansard and A. Colin,
Soft Matter \textbf{8}, 4025 (2012).

\bibitem{AT01}
I.S. Aranson and L.S. Tsimring,
Phys. Rev. E {\bf 64}, 020301(R)  (2001).

\bibitem{AT02}
I.S. Aranson and L.S. Tsimring,
Phys. Rev. E {\bf 65}, 061303 (2002).

\bibitem{VTA03}
D. Volfson, L.S. Tsimring and I.S. Aranson,
Phys. Rev. E {\bf 68}, 021301 (2003).

\bibitem{ATMC08} 
I.S. Aranson, L.S. Tsimring, F. Malloggi and E. Cl\'ement,
Phys.Rev.E \textbf{78}, 031303 (2008).

\bibitem{O08}
P. D. Olmsted,
Rheol. Acta \textbf{47}, 283 (2008).

\bibitem{TBKCCA15}
M. Trulson, M. Bouzid, J. Kurchan, E. Cl\'ement, P. Claudin, and B. Andreotti,
Europhys. Lett. \textbf{111}, 18001 (2015). 

\bibitem{EAN96}
M.D. Ediger, C.A. Angell and S.R. Nagel,
J. Phys. Chem. \textbf{100}, 13200 (1996). 

\bibitem{DS01} 
P. Debenedetti and F. Stillinger,
Nature \textbf{410}, 259 (2001).

\bibitem{MKK09}
R. Mari, F. Krzakala and J. Kurchan,
Phys. Rev. Lett. \textbf{103}, 025701 (2009).

\bibitem{OT07}
P. Olsson and S. Teitel,
Phys. Rev. Lett. \textbf{99}, 178001(2007).

\bibitem{PvM85} 
P.N. Pusey and W. van Megen, 
Nature \textbf{320}, 340 (1985).

\bibitem{MW95}
T.G. Mason and D.A. Weitz,
Phys. Rev. Lett. \textbf{75}, 2770 (1995).

\bibitem{HW95}
G.L. Hunter and E.R. Weeks,
Rep. Prog. Phys. \textbf{75}, 066501 (2012).

\bibitem{MW12}
J. Mewis and N. Wagner,
\textit{Colloidal Suspension Rheology},
Cambridge University Press, New York (2012).

\bibitem{CBML03}
M. Clo\^\i tre, R. Borrega, F. Monit and L. Leibler,
Phys. Rev. Lett. \textbf{90}, 068303 (2003) .

\bibitem{F98}
D.S. Fisher,
Physics Reports \textbf{301}, 113 (1998).

\bibitem{TLB06}
A. Tanguy, F. Leonforte and J.-L. Barrat,
Eur. Phys. J E \textbf{20}, 355 (2006).

\bibitem{HKLP10}
H.G.E. Hentschel, S. Karmakar, E. Lerner and I. Procaccia,
Phys. Rev. Lett. \textbf{104}, 025501 (2010).

\bibitem{AFP13}
B. Andreotti, Y. Forterre and O. Pouliquen,
\textit{Granular Media: Between Fluid and Solid},
Cambridge University Press (2013).

\bibitem{LN98}
A.J. Liu and S.R. Nagel,
Nature \textbf{396}, 21 (1998).

\bibitem{GDRMidi} 
GDR MiDI,
Eur. Phys. J. E. \textbf{14}, 341 (2004).

\bibitem{CEPRC05}
F. da Cruz, S. Emam, M. Prochnow, J.N. Roux and F. Chevoir,
Phys. Rev. E \textbf{72}, 021309 (2005).

\bibitem{TBCA13}
M. Trulsson, M. Bouzid, P. Claudin and B. Andreotti
Europhys. Lett. \textbf{103}, 38002 (2013). 

\bibitem{P99} 
O. Pouliquen,
Phys. Fluids. \textbf{11}, 542 (1999).

\bibitem{JFP06}
P. Jop, Y. Forterre and O. Pouliquen,
Nature. \textbf{441}, 727 (2006).

\bibitem{BGP11}
F. Boyer, E. Guazzelli and O. Pouliquen,
Phys. Rev. Lett. \textbf{107}, 188301 (2011).

\bibitem{TAC12}
M. Trulsson, B. Andreotti and P. Claudin,
Phys. Rev. Lett. \textbf{109}, 118305 (2012).

\bibitem{BTCCA13}
M. Bouzid, M. Trulsson, P. Claudin, E. Cl\'ement and B. Andreotti,
Phys. Rev. Lett. \textbf{111}, 238301 (2013). 

\bibitem{KTMvH10}
G. Katgert, B.P. Tighe, M.E. M\"obius and M. van Hecke,
Europhys. Lett. \textbf{90}, 54002 (2010).

\bibitem{JS83} 
J.T. Jenkins and S.B. Savage,
J. Fluid Mech. \textbf{130}, 187 (1983).

\bibitem{KINN01} 
T.S. Komatsu, S. Inagaki, N. Nakagawa and S. Nasuno,
Phys. Rev. Lett. \textbf{86}, 1757 (2001).

\bibitem{TRVLPJD03}
N. Taberlet, P. Richard, A. Valance, W. Losert, J.M. Pasini, J.T. Jenkins and R. Delannay,
Phys. Rev. Lett. {\bf 91}, 264301 (2003)

\bibitem{TRD08}
N. Taberlet, P. Richard and R. Delannay,
Computers and Mathematics with Applications \textbf{55}, 230 (2008).

\bibitem{NDBC11}
V.B. Nguyen, T. Darnige, A. Bruand and E. Cl\'ement,
Phys. Rev. Lett. \textbf{107}, 138303 (2011).

\bibitem{MAC15}
F. Malloggi, B. Andreotti and E. Cl\'ement,
Phys. Rev. E \textbf{91}, 052202 (2015). 

\bibitem{DLDA06}
S. Deboeuf, E. Lajeunesse, O. Dauchot and B. Andreotti,
Phys. Rev. Lett. \textbf{97}, 158303 (2006).

\bibitem{NZBWH10}
K. Nichol, A. Zanin, R. Bastien, E. Wandersman and M. van Hecke,
Phys. Rev. Lett. \textbf{104}, 078302 (2010).

\bibitem{RFP11}
K.A. Reddy, Y. Forterre and O. Pouliquen,
Phys. Rev. Lett. \textbf{106}, 108301 (2011).

\bibitem{WH14}
E. Wandersman, M. Van Hecke,
Europhys. Lett. \textbf{105}, 24002 (2014).

\bibitem{HDKC11}
R. Harich, T. Darnige, E. Kolb and E. Cl\'ement,
Europhys. Lett \textbf{96}, 54003 (2011).

\bibitem {MV87}
H.B. M\"uŸhlhaus and I. Vardoulakis, 
G\'eotechnique \textbf{37}, 271 (1987)

\bibitem{DAL01}
C. Derec, A. Ajdari and F. Lequeux,
Eur. Phys. J. E \textbf{4}, 355 (2001).

\bibitem{MLAC06}
F. Malloggi, J. Lanuza, B. Andreotti and E. Cl\'ement,
Europhys. Lett. \textbf{75}, 825 (2006).

\bibitem{AMC06}
I.S. Aranson, F. Malloggi, and E. Cl\'ement,
Phys. Rev. E \textbf{73}, 050302(R) (2006).

\bibitem{CMAA07}
E. Cl\'ement, F. Malloggi, B. Andreotti, I.S. Aranson,
Granular Matter \textbf{10}, 3 (2007).

\bibitem{KK12}
K. Kamrin and G. Koval,
Phys. Rev. Lett. \textbf{108}, 178301 (2012).

\bibitem{PF09} 
O. Pouliquen and Y. Forterre,
Phil. Trans. R. Soc. A. \textbf{367}, 5091 (2009).

\bibitem{PhDBouzid14}
M. Bouzid, 
PhD Thesis, \textit{Comportement rh\'eologiques et effets non-locaux dans les \'ecoulements granulaires denses.}
Universit\'e Paris Diderot (2014).

\bibitem{JS77}
M. Johnson and D. Segalman,
J. Non-Newt. Fluid Mech. \textbf{2}, 255 (1977). 

\bibitem{ORL00}
P. D. Olmsted, O. Radulescu and C.-Y. D. Lu
Jou. Rheol. \textbf{44}, 257 (2000).

\bibitem{MG07}
P. Marmottant, F. Graner
Eur. Phys. J. E. \textbf{23}, 337 (2007).

\bibitem{ANBCC12}
A. Amon, V. B. Nguyen, A. Bruand, J. Crassous, and E. Cl\'ement,
Phys. Rev. Lett. \textbf{108}, 135502 (2012).

\bibitem{CDW15}
D. Chen, K.W. Desmond and E.R. Weeks,
Phys. Rev. E \textbf{91}, 062306 (2015).

\bibitem{DW15}
K.W. Desmond and E.R. Weeks,
Phys. Rev. Lett. \textbf{115}, 098302 (2015).

\bibitem{FL98}
M.L. Falk and J.S. Langer
Phys. Rev. E \textbf{57}, 7192 (1998).

\bibitem{HL98}
P. H\'ebraud and F. Lequeux,
Phys. Rev. Lett. \textbf{81}, 2934 (1998).

\bibitem{MBC14}
V. Mansard, L. Bocquet and A. Colin,
Soft Matter \textbf{10}, 6984 (2014).

\bibitem{LC09}  
A. Lema\^\i tre and C. Caroli,
Phys. Rev. Lett.  \textbf{103}, 065501 (2009).

\bibitem{HK13}
D.L. Henann and K. Kamrin,
Proc. Natl. Acad. Sci. USA \textbf{110}, 6730 (2013).

\bibitem{GG01}
B.J. Glasser and I. Goldhirsch,
Phys. Fluids \textbf{13}, 407 (2001).

\bibitem{GG02}
C. Goldenberg and I. Goldhirsch,
Phys. Rev. Lett. \textbf{89}, 084302 (2002).

\bibitem{HK14}
D.L. Henann and K. Kamrin,
Phys. Rev. Lett. \textbf{113}, 178001 (2014).

\bibitem{BSSPBST15}
R. Benzi, M. Sbragaglia, A. Scagliarini, P. Perlekar, M. Bernaschi, S. Succi and F. Toschi,
Soft Matter \textbf{11}, 1271 (2015).

\bibitem{DBZU11}
K.A. Dahmen, Y. Ben-Zion, and J.T. Uhl,
Nature Phys. \textbf{7}, 554 (2011).

\bibitem{BTCCA15}
M. Bouzid, M. Trulsson, P. Claudin, E. Cl\'ement and B. Andreotti,
Europhys. Lett. \textbf{109}, 24002 (2015).

\bibitem{RTACS15}
E. Rojas, M. Trulsson, B. Andreotti, E. Cl\'ement and R. Soto,
Europhys. Lett. \textbf{109}, 64002 (2015).

\bibitem{LBAMcNC14}
A. Le Bouil, A. Amon, S. McNamara, and J. Crassous,
Phys. Rev. Lett. \textbf{112}, 246001 (2014).

\bibitem{W09}
M. Wyart,
Europhys. Lett. \textbf{85}, 24003 (2009).

\end{thebibliography}
%

\end{document}